# Losses of thulium atoms from optical dipole traps operating at 532 and 1064 nm

V.V. Tsyganok [1], D.A. Pershin [1,2], V.A. Khlebnikov [1], D.A. Kumpilov[1,3], I.A. Pyrkh[1,4], A.E. Rudnev[1,3], E.A. Fedotova[1,3], D.V. Gaifudinov[1,3], I.S. Cojocaru[1,2], K.A. Khoruzhii[1,3], P.A. Aksentsev[1,4], A.K. Zykova[1] and A.V. Akimov[1,2,5]

[1]*Russian Quantum Center, Bolshoy Boulevard 30, building 1, Skolkovo, 143025, Russia*

[2]*PN Lebedev Institute RAS, Leninsky Prospekt 53, Moscow, 119991, Russia*

[3]*Moscow Institute of Physics and Technology, Institutskii pereulok 9, Dolgoprudny, Moscow Region 141701, Russia*

[4]*Bauman Moscow State Technical University, 2-nd Baumanskaya, 5, Moscow, 105005, Russia*

[5]*National University of Science and Technology MISIS, Leninsky Prospekt 4, Moscow, 119049, Russia*

email: *a.akimov@rqc.ru*

Recently thulium has been condensed to Bose-Einstein condensate. Machine learning was used to avoid a detailed study of all obstacles making cooling difficult. This paper analyses the atomic loss mechanism for the 532 nm optical trap, used in the Bose-condensation experiment, and compares it with the alternative and more traditional micron-range optical dipole trap. We also measured the scalar and tensor polarizability of thulium at 1064 nm and was found to be $167 \pm 25$ a.u. ( $275 \pm 41 \times 10^{-41} F \cdot m^2$ ) and $-4 \pm 1$ a.u. ( $7 \pm 2 \times 10^{-41} F \cdot m^2$ ).

## I. INTRODUCTION

The field of ultracold atoms is actively developing due to a number of interesting applications and research information available with such systems. These include timekeeping [1–3] and search for drift of fundamental constants [4,5], sensing of gravitational [6,7] and magnetic fields [8,9], measurements of rotation [10], and quantum simulations [11]. Thulium may be particularly interesting for the realization of quantum simulations due to its large magnetic moment in the ground state (4 Bohr magnetons), facilitating long-term interactions and easily accessible low-field Feshbach resonances [12,13] enabling control over short-term interactions. Recently, thulium has been condensed using evaporative cooling [14]: the machine learning technique has been used to optimize the evaporative process. The cooling was performed in an optical dipole trap (ODT) with a wavelength of 532 nm. This approach allowed the cooling down of thulium atoms without any analysis of the main loss mechanisms. On the other hand, knowledge of possible loss mechanisms may help further research with a cold atomic cloud or condensate. The trap loss mechanisms could be studied by observing the dynamics of a trapped atomic sample.

This study analyses the loss mechanism of atoms from the 532 nm dipole trap and indicates good congruency between the loss rate and radiative loss model. In addition, the polarizability of thulium around 1064 nm was also measured and, although calculated previously [15], to our best knowledge it has not been experimentally verified. The losses of the 1064 nm ODT were also measured. In this case, the trap loss rate was considerably higher than was predicted by the radiative loss model but the absolute value was smaller than in the case of the 532 nm dipole trap.

## II. LOSSES FOR THE 532 ODT

The details of the experimental setup could be found elsewhere [16–20]. The atoms were precooled in Zeeman slower and 2D optical molasses, operating at strong transition $4f^{13}(^2F^0)6s^2 \to 4f^{12}(^3H_5)5d_{3/2}6s^2$ with a wavelength of 410.6 nm and a natural width of $\Gamma = 2\pi\gamma = 2\pi \cdot 10.5$ MHz. Then atoms were trapped into a magneto-optical trap operating on weaker transition $4f^{13}(^2F^o)6s^2 \to 4f^{12}(^3H_6)5d_{5/2}6s^2$ with a wavelength of 530.7 nm and a natural width of $\Gamma = 2\pi\gamma = 2\pi \cdot 345.5$ kHz [21]. By using a large detuning of the MOT light along with reduction of its intensity [19], atoms were polarized to the lowest magnetic sublevel of the ground state $|F=4; m_F=-4\rangle$ [22–25] and cooled to a temperature of around 13 µK. Then atoms were loaded into the optical dipole trap (see Figure 1a, b) [16]. The trap was formed by a laser beam with a wavelength of 532.07 nm focused on the beam waist of 15.8 µm and 25.7 µm in the $z_g$ and $y_g$ directions, respectively. The waist along $x_g$ direction was about 2 mm. Polarization of the beam was linear with angle 90 degrees to z direction. The trap was swept using an acousto-optic modulator to increase overlap with the magneto-optical trap so that $w_y^* = 170$ μm [14]. Once loaded into the ODT, the atomic cloud was evaporatively cooled using the forced evaporation cooling sequence, described in [14] (see Figure 2a) with a constant Oz-oriented magnetic field of $-3.91\,\text{G}$. During evaporative cooling, the second, "vertical" ODT beam circularly polarized $\varepsilon = |u^* \times u| = -0.95$ (here $u$ – normalized Jones vector) with a waist 90 $\mu$m was turned on to increase the confinement of an atomic cloud and to increase the collisional rate during the evaporation (see Figure 1b).

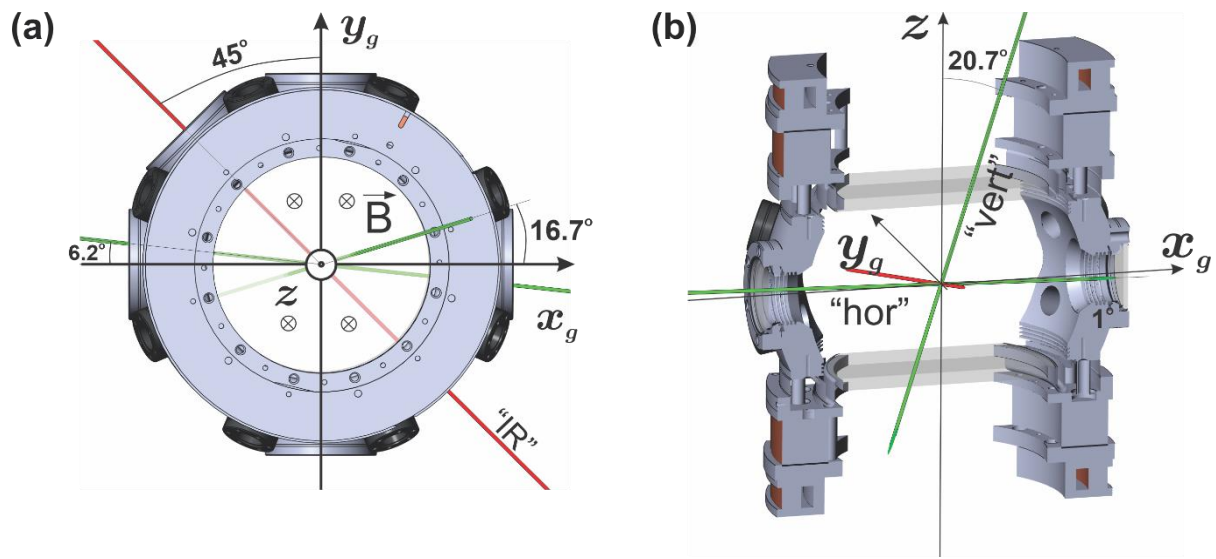

*Figure 1. Configuration of the optical beams forming the dipole trap. (a) top view, the red beam stands for the "IR" 1064 nm laser, green for 532 nm laser ODT beams, B stands for the magnetic field; (b) side view with "vertical"("vert") and "horizontal"("hor") ODT beams marked.*

To gain insight into the mechanisms of atomic losses from the trap, it is important to keep the trap depth to the atomic temperature ratio as high as $\eta > 6:1$ [26,27]. In this case, the self-evaporation of the atomic sample is significantly suppressed. This was achieved by a sudden rise (200 ms ramp in Figure 2a around 3.1 s) of the power in both ODT beams ("Hor." and "Vert." in Figure 2a and Figure 1b) thus setting frequencies of the ODT to be $(v_x, v_y, v_z) = (325, 85, 442)$ Hz (see [28] for ODT frequency measurements). As a result, the temperature of atoms after 150 ms of the thermalization time interval was found to be 420 nK, while the depth of the trap was 6.6 µK. It should be noted that the thermalization time was measured to be 22 ms [29].

Once thermalization was complete, the long-term loss of the atoms, as well as their temperature, were both measured for various delays $\Delta t$ (see Figure 2a). To detect the number of atoms, the absorption technique, described in [14], was used. The temperature was measured as

$$T = \frac{m_{Th}}{2k_B} \frac{\sigma_i(t)^2}{1/(2\pi v_i)^2 + t^2}, \quad (1)$$

where $m_{Th}$ – thulium atomic mass, $k_B$ – Boltzmann constant, $\sigma_i(t)$ – atomic cloud size at the level $e^{-1}$ at $t$ seconds of the ballistic expansion (blue dots in Figure 2c) and with the standard ballistic method (orange dots). Figure 2b plots the number $N$ of detected atoms in the trap versus time. It is clearly seen that in the deep ODT, the decay of $N$ is exponential $N = N_0 e^{-\Delta t/\tau}$ and the decay constant was found to be $\tau = 5.2 \pm 0.1$ s. This decay is accompanied by the slow heating of the remaining atoms in the trap, as can be seen in Figure 2c. The heating follows the linear model $T = T_0 + \alpha \Delta t$ with $\alpha = 174 \pm 18$ nK/s. The mean density $n$ of atoms in the harmonic trap (Figure 2d) is a function of both the number of atoms $N$ and the temperature since:

$$n = N(2\pi)^3 (v_x v_y v_z) \left(\frac{m_{th}}{2\pi k_B T}\right)^{3/2} \frac{1}{\sqrt{8}} \quad (2)$$

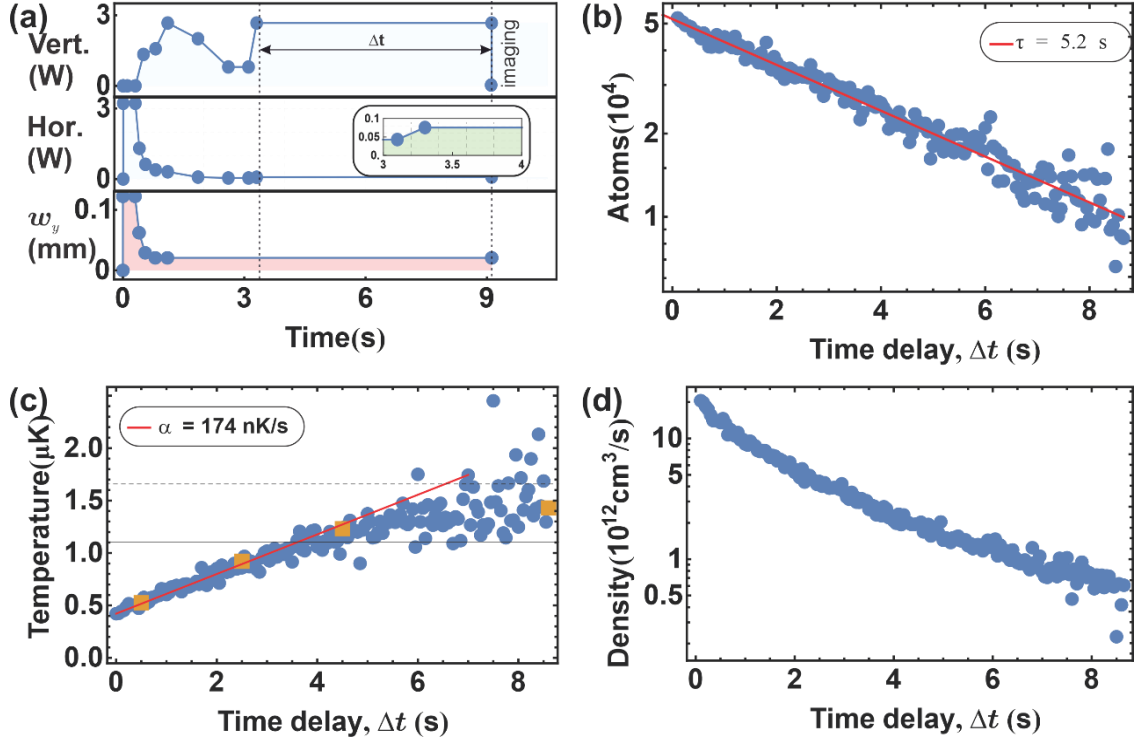

*Figure 2. a) Evaporation sequence. Vert. stand for power of vertical ODT beam, Hor. -- for power of horizontal ODT beam, $w_x$ -- for the horizontal beam waist. The initial state of the atoms is prepared into the crossed ODT at 532 nm for the first 3.1 s. After 3.1 s, the depth of the crossed ODT increases by 1.8 times. b), c) and d) – evolution of the number of atoms, temperature, and the mean density of the atomic ensemble in the ODT after increasing depth. The red line in b) is a linear fit of the data. Solid and dashed black lines represent temperature levels for which the ratio between the depth of the ODT and temperature becomes 6 and 4 correspondingly. Only data before the black line (or 4 s) is used in data interpretation. Orange dots correspond to the temperature that was found from the ballistic expansion of the atomic ensemble. Blue dots correspond to a rough estimate of the temperature found from the size of the atomic cloud. The red line in c) is a linear fit of the data before 4 s.*

Different mechanisms of atomic losses lead to different decay characters [21,30–35]. The following loss mechanisms are usually considered: single atom losses due to elastic collisions with a hot background gas and/or an evaporation process that is an escape of an atom from the trap after energy and momentum redistribution along elastic collision with another trapped atom; three-body inelastic losses and there is also a possibility of photon absorption from ODT beams. Photon absorption is a heating-related mechanism of losses, while collisions with buffer gases contribute negligibly to heating, while 3-body recombination contributes to both losses and heating. Overall, this process can be summarized [36] in the form of the following equations for the number of atoms $N$ and temperature $T$:

$$\frac{dN}{dt} = -\alpha N - \gamma \frac{N^3}{T^3},$$
$$\frac{dT}{dt} = \frac{1}{3} T_{rec} \Gamma_{SC} + \gamma \frac{N^2}{T^3} \frac{(T+T_h)}{3},$$
(3)

where $\alpha$, $\gamma$ are one- and three-body loss rates, $T_{rec} = 2E_{recoil}/k_B$ – recoil temperature, . $E_{rec} = (\hbar k)^2 / 2m_{th}$. – recoil energy, $\Gamma_{SC}$ – off-resonant photon scattering rate, $T_h$ – recombination constant, corresponding to heating energy per lost atom in a three-body recombination process that was seen in system with high density and large scattering length [34].

In the discussed case, evaporation cooling is prevented by keeping the depth-to-temperature ratio $\eta > 6$, and three-body losses are suppressed by setting the magnetic field far from any Feshbach resonance [13].

The observed losses are described by a single exponential function (see Figure 2b), and heating is linear with time (see Figure 2c) thus indeed making it possible to exclude the contribution of 3-body processes. This means that the only important process must be the collisions with a buffer gas (see Figure 2b) and heating via ODT light absorption, which is most likely caused by the closeness of the 532 nm ODT light to 532.7 nm atomic transition. In this case [13], the equation (3) could be rewritten as:

$$\dot{N} = \frac{N}{\tau},$$
$$\dot{T} = \frac{1}{3} T_{rec} \Gamma_{SC},$$
(4)

where $\tau = 1/\alpha$ – the lifetime of the ODT.

The heating rate can be estimated from the previously measured [28] imaginary part of the polarizability as [16,37,38]:

$$\Gamma_{SC} = \frac{4\pi a_B^3}{\hbar c} I(r) \left[ \kappa_{sc} + |u^* \times u| \cos\theta_k \frac{m_F}{2F} \kappa_{vec} - \frac{3m_F^2 - F(F+1)}{F(2F-1)} \frac{f(\theta_k, \theta_p, |u^* \times u|)}{2} \kappa_t \right],$$
$$f(\theta_k, \theta_p, |u^* \times u|) = 1 - \frac{3}{2} (\sin\theta_k)^2 \left(1 + \sqrt{1 - |u^* \times u|^2} \cos(2\theta_p)\right)$$
(5)

where $u$ is a Jones vector, $\theta_p = \angle(\vec{E}, \vec{B})$ and $\theta_k = \angle(\vec{k}, \vec{B})$, $\vec{k}$ is a wave vector, $\kappa_{sc}, \kappa_{sc}, \kappa_{sc}$ are imaginary parts of scalar, vector, and tensor polarizability.

Given the experimentally measured powers of two beams 76 mW (horizontal, see Figure 1) and 2.67 W (vertical), $\theta_{k,hor} = 89 \pm 3°$, $\theta_{P,hor} = 90.0 \pm 0.1°$ as well as $\theta_{k,vert} = 21 \pm 2°$, $\theta_{P,hor} = 79.3 \pm 0.1°$, $\varepsilon_{hor} = 0$ and $\varepsilon_{vert} = -0.95$, that is:

$$\Gamma_{SC} = \Gamma_{SC,hor}(P_{hor}, \varepsilon_{hor}, \theta_{k,hor}, \theta_{P,hor}) + \Gamma_{SC,vert}(P_{vert}, \varepsilon_{vert}, \theta_{k,vert}, \theta_{P,vert}) \simeq 1.5 \text{ Hz} . \quad (6)$$

The measured heating rate of $\alpha = 174 \pm 18$ nK/s corresponding to $\Gamma_{SC} = 1.3 \pm 0.1$ Hz is indeed close to its theoretical estimation. While these losses do not prevent evaporative cooling down to Bose-Einstein condensation, it is interesting to compare those with ones in a more traditional ODT with a wavelength of 1064 µm, far away from all atomic transitions.

## III. POLARISABILITY OF THULIUM NEAR THE 1064 NM

The differential dynamic polarizability of thulium atoms at a wavelength of 1064 nm was measured experimentally [15], but absolute numbers for the polarizability have not been experimentally measured yet. Knowledge of the full value of atomic polarizability, as was shown above, is crucial for the analysis of the trapped atomic cloud decay. Therefore, the polarizability of thulium around 1064 nm was measured before the measurements of ODT lifetime for this wavelength. To study the interaction of the atomic ensemble of thulium with a 1064 nm laser beam, one of the windows of the vacuum chamber was replaced by another one having antireflection coating at 1064 nm. The other side of the chamber had a long metallic sleeve of about 51.5 cm. The long distance along with a rather tight focusing of the beam ensures the absence of noticeable reflections and thus the window at the end of the long metallic sleeve was kept untouched.

Spherically asymmetrical atoms, including lanthanides, have a nonzero orbital angular momentum in the ground state, which leads to a contribution of the tensor and vector parts to the real part of dynamic polarizability $\alpha_{total}$ in the ground state [16,37,38]:

$$\alpha_{total} = \alpha_{sc} + |u^* \times u| \cos\theta_k \frac{m_F}{2F} \alpha_{vec}(\omega) - \\ -\frac{3m_F^2 - F(F+1)}{F(2F-1)} \frac{f(\theta_k, \theta_p, |u^* \times u|)}{2} \alpha_t(\omega) \quad . \tag{7}$$

To calculate atomic polarizability at a wavelength of 1064 nm, the sum contribution of all known polarizabilities from the National Institute of Standards' and Technology database [39] and formulas (4,5,6) from [40]. Accordingly, the real and imaginary parts of the polarizabilities were calculated to be:

$$\begin{aligned}
\alpha_{SC} &= 159.6 \ a.u. \\
\alpha_{TEN} &= -3.2 \ a.u. \\
\alpha_{VEC} &= -0.2 \ a.u. \\
\kappa_{SC} &= 16.6 \times 10^{-7} \ a.u. \\
\kappa_{TEN} &= -2.6 \times 10^{-7} \ a.u. \\
\kappa_{VEC} &= -0.4 \times 10^{-7} \ a.u.
\end{aligned} \tag{8}$$

To find experimental values for polarizabilities one need to express polarizability via experimentally measured parameters. It is convenient to use for this purpose trap frequencies $\nu_i$, where $i \in \{x, y, z\}$ in the coordinate system, shown in Figure 3b. The real part of polarizability $\alpha_{tot}$ can be expressed via trap frequencies as [19]:

$$\alpha_{tot} = \pi^3 \varepsilon_0 c \frac{\nu_y^2 w_y^3 w_z m_{Th}}{P} = \pi^3 \varepsilon_0 c \frac{\nu_z^2 w_z^3 w_y m_{Th}}{P} \quad . \tag{9}$$

where $m_{Th}$ is atomic mass, $w_x$ is Rayleigh length and $w_y, w_z$ are beam waists at the $e^{-2}$ level, $P$ is a beam power.

## IV. EXPERIMENTAL MEASUREMENTS NEAR THE 1064 NM

The preparation of the thulium ensemble in a 1064 nm ODT is the same as in a 532 nm ODT. The atomic cloud of $^{169}$Tm was initially cooled down with MOT in a polarized configuration [19] to prepare atoms in $|J = 7/2, F = 4, m_F = -4\rangle$. Then atoms were loaded into a scanning 1064 nm ODT with an aspect ratio $\rho = w_y/w_z = 5.7$ and vertical (Oz-axis) and "holding" magnetic field of $-3.91 \ \text{G}$ set far from Feshbach resonances [13]. Typically, the swept ODT has about $8.7 \times 10^6$ atoms with a temperature of about $23 \ \mu\text{K}$. After switching off the sweeping of the ODT and evaporative cooling for 2.5 s, the remaining number of atoms was about $1.1 \times 10^6$ with a temperature of around 1 µK. To achieve a larger depth-to-temperature ratio, the power $P$ of the ODT beam was enlarged linearly within $t_r = 1.5$ s to different target values $P_F$ (see Figure 3a).

Beam waist is a parameter with enters into expression (9) in 4th power (there is cube of a waste in $y$ direction and one more waist factor in $z$ direction). Thus, relative uncertainty in measurements of the beam waist leads to approximately 4-fold increase in relative uncertainty for polarizability. To reduce this uncertainty and minimize the error bar for the polarizability, the waist $w_y$ was measured in a sweeping regime with big $\rho$ for several values of target waist $w_{yF}$ (see Figure 3a) [19].

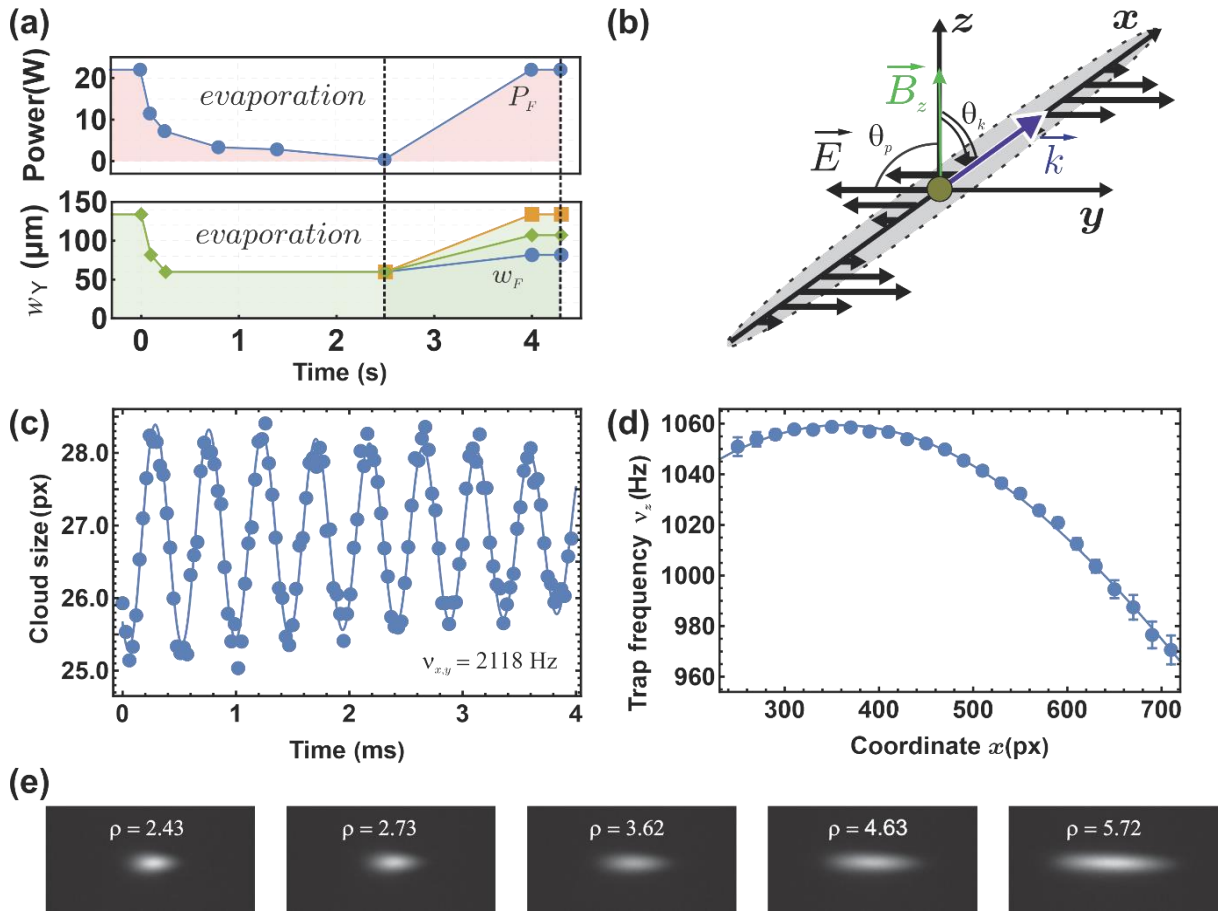

Figure 3. a) Experimental sequence of evaporation in a 1064 nm ODT and trap frequency measurements. b) The geometry of the experiment with the polarizability of thulium atoms: a polarized atomic cloud ($|F = 4; m_F = -4\rangle$) oriented along the magnetic field $\vec{B}$ in a single ODT beam along the x-axis with linear ODT light polarization along the y-axis. c) Typical cloud size oscillations with time and their fit with expression (12) for a fraction of the trapped atomic cloud near the position

*x=370 px from panel (d). d) Dependence of the fitted frequency versus the coordinate x along the horizontal beam when it is imaged in xz plane. The trap frequency is half of the cloud size frequency. e) Pictures of the beam at its waist for various sweeping amplitudes.*

A CCD camera (Thorlabs DCU223M-GL) was used to measure the waist size $w_y$. The ODT beam was deflected on a mirror aside from the vacuum volume. The camera was set at the focus of the ODT beam. The resulting beam profile (see Figure 3e) was approximated by Gaussian. The pixel size was taken as 4.65 µm from the camera specification.

Then the deflecting mirror was removed. The frequencies of the trap in both $z$ and $y$ directions were measured for each known waist size $w_y$, as in our previous study, by fitting oscillation frequency and finding its maximum over spatial coordinates [28] (see Figure 3c,d). Considering that the beam waist $w_{yF}$ for big $\rho$ is well known (see Figure 3a and the inset on Figure 4c), another waist at the z-axis can be found as:

$$w_z = w_{yF} \frac{\nu_z}{\nu_y}, \qquad (10)$$

and it is $w_z = 25.3 \pm 0.9$ µm. In its turn, $w_Y$ without a sweeping regime can be found from the relation of the slope of frequency versus the root of power (see Figure 4b) and is $w_y = 57.6 \pm 2$ µm. The resulting aspect ratio was compared with waists from the gauss fit of the ODT beam at focus (see Figure 4c).

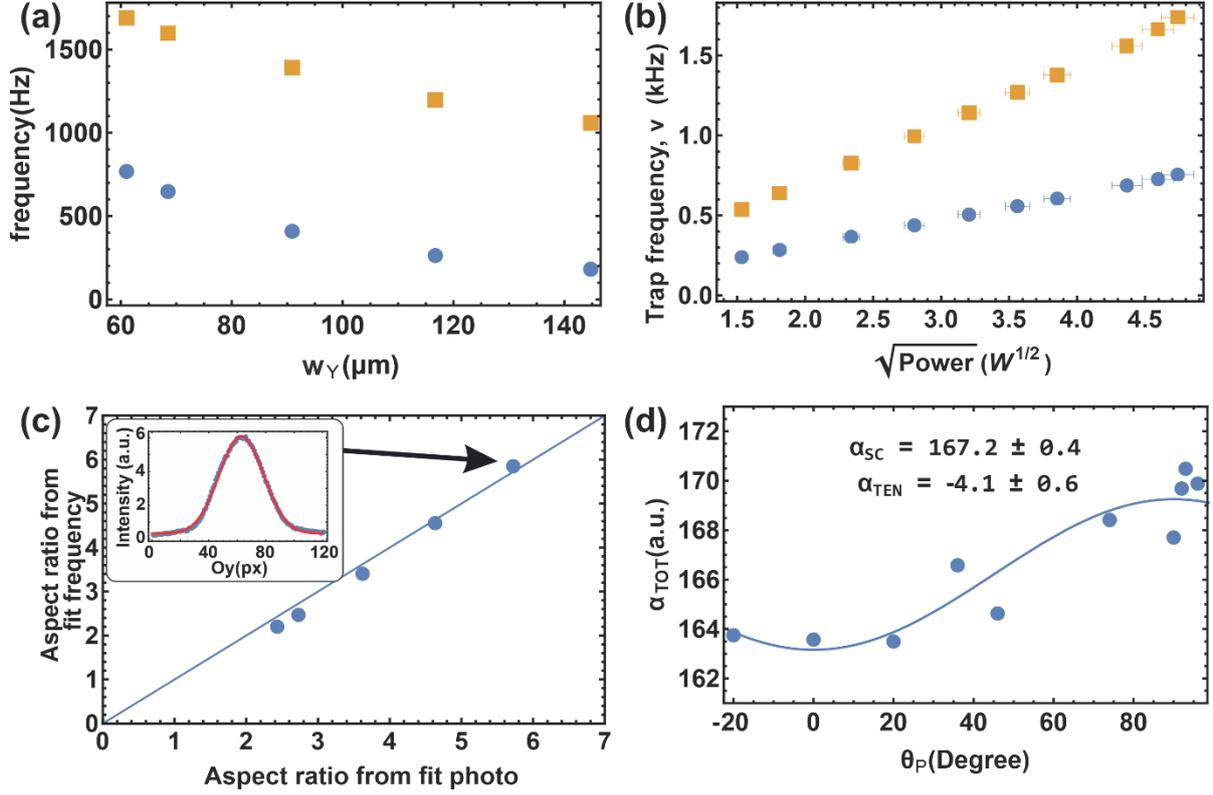

*Figure 4. a) ODT frequencies in y- (blue dots) and z- (orange squares) directions versus the waist $w_y$. b) ODT frequencies in y- (blue dots) and z- (orange squares) directions versus the root of power. c) The aspect ratio from the frequency fit compared to the aspect ratio from the photo of the ODT beam. The blue line is 45 degrees tilted. Inset – the intensity distribution of the ODT beam in the y-direction in the sweeping regime with AR of 5.72 with its gaussian fit (red line). d) Total polarizability of thulium atoms versus the angle $\theta_P$ and the fit by expression* (11).

As can be seen from Eq. (7), when $\theta_k = 90°$ or $\varepsilon = |u^* \times u| = 0$, the vector term becomes zero, and total polarizability becomes:

$$\alpha_{tot} = \alpha_{SC} + \frac{3m_F^2 - F(F+1)}{F(2F-1)} \times \frac{3\cos^2\theta_P - 1}{2} \alpha_{tens}. \tag{11}$$

To measure tensor and scalar polarizability, a linearly polarized ODT with $\theta_k = 90°$ was used. The $\theta_P$ angle was varied and the total polarizability (see Figure 4d) was extracted from trap frequency measurements using:

$$w_z = Ampl \cos[2\pi v_z t + \phi] + w_{z0}, \tag{12}$$

where $Ampl, \nu_z, \phi, w_{z0}$ – parameters of the fit. During these experiments, the magnetic field was kept constant having a value $-3.91\,\text{G}$ and was oriented along the $z$-axis. The orientation of the ODT polarization was controlled by a half-wave plate and was measured by a polarization beam splitter placed after the vacuum chamber. Using (11) and eq. (9), polarizabilities were found to be:

$$\begin{aligned} \alpha_{SC} &= 167.2 \pm 0.4_{STAT} \pm 24.7_{SYS} \text{ a.u.} \\ \alpha_{TEN} &= -4.1 \pm 0.6_{STAT} \pm 0.6_{SYS} \text{ a.u.} \end{aligned} \quad (13)$$

These values are consistent with the theoretical predictions in Eq. (8).

Theoretically, the vector part of polarizability around 1064 nm is expected to be very small (see eq. (8)). To verify this, the circularly polarized light beam with $|u^* \times u| = 0.96$ was used in the ODT. While preparation of atomic ensemble is still done at $-3.91\,\text{G}$, rotation of the magnetic field, necessary in these experiments requires use of smaller field due to technical limitation of the equipment. Therefore, after initial preparation the magnetic field was reduce in magnitude to $2.9\,\text{G}$, still far from Feshbach resonance and then the field was rotated to be parallel to the x-axis. In this configuration, the contribution of vector polarizability into a total one is maximized, see (5), (7). To exclude the role of remaining tensor polarizability, the measurements for total polarizability were done for two magnetic field orientations: $+2.9\,\text{G}$ and $-2.9\,\text{G}$. The necessity to flip the current direction limits the available magnetic field range. Total polarizabilities are the same in these two cases and equal to $166 \pm 25$ a.u., which is consistent with the expectation of very small vector polarizability.

## V. HEATING IN A 1064 NM ODT

The heating rate in the 1064 nm ODT was also investigated. The depth of the ODT was about $350\,\mu\text{K}$; therefore, the depth-to-temperature ratio was $\eta = 13:1$, and the evaporation process was reliably suppressed. The ODT polarization was kept linear in these experiments. Atomic loss for $\theta_p = 90°$ is shown in Figure 5a; it has single exponential behavior with time and was fitted with eq. (4). The temperature of the cloud has linear behavior in this experiment (Figure 5b) with $\alpha = 1-2.3\,\mu K/s$. Figure 5c,d presents the heating rate and lifetime of the ODT versus the angle $\theta_P$.

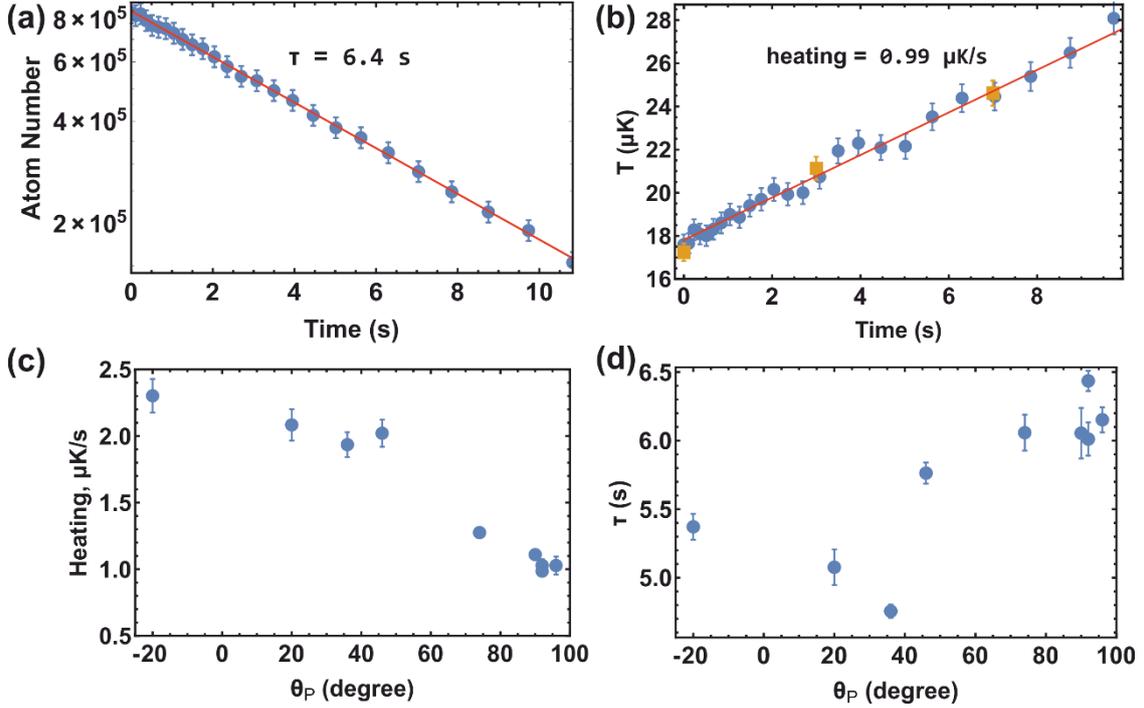

*Figure 5. a) Atom loss in a deep 1064 nm ODT with $\theta_P = 90°$. The ratio of the potential depth to temperature corresponds to 21 at the start and 13 at the end of measuring. The red line is exponential fit. b) Heating of the atomic ensemble during the holding time in a deep 1064 nm ODT. Blue dots correspond to a rough estimate of the temperature found from the size of the atomic cloud. Orange dots correspond to the temperature that was found from the ballistic expansion of the atomic ensemble. The red line is a linear fit of the experimental data. c) Dependence of the off-resonance photon scattering rate versus $\theta_P$ – the angle between the light polarization plane and quantization axis (external magnetic field). d) Dependence of the lifetime of the ODT versus $\theta_P$*

According to (8) and (4), heating for $\theta_P = 90°$, $\varepsilon = 0$ and $\theta_k = 90°$ in this experiment should be about 31 nK/s. Thus, the heating rate, in this case, cannot be explained by the absorption of photons from the trap light. At the same time, since both the heating rate is linear and losses are well described by the single exponential function, the 3-body mechanism still has to be excluded. Therefore, there is a not yet identified heating mechanism, such as small beam vibrations or laser power fluctuations [41–43], probably due to technical limitations. These mechanisms were not investigated in detail.

## VI. CONCLUSION

Using deep optical dipole trap the loss mechanism form optical dipole trap, operating at 532 nm was analyzed. The heating due to absorption of photons from optical dipole trap mechanism is adequately describes observed experimental losses from optical dipole trap and therefore was indicated as the main loss mechanism. To the contrary, in the case of 1064 nm dipole trap the losses from the trap happened to be considerably higher than that predicted by the same radiative model and overall improvement in the trap lifetime was found to be rather moderate, improving lifetime from 5.2 to 6.4 seconds. This lifetime is nevertheless sufficient for the future experiments planned at the laboratory. The polarizability of thulium atom at 1064 nm was also experimentally investigated. The values of polarizability were found to be $167 \pm 25$ a.u. ($275 \pm 41 \times 10^{-41} \text{F} \cdot \text{m}^2$) and $-4 \pm 1$ a.u. ($7 \pm 2 \times 10^{-41} \text{F} \cdot \text{m}^2$) and are in a good agreement with theoretically predicted values for scalar $\alpha_{SC} = 159.6$ $a.u.$, and tensor $\alpha_{TEN} = -3.2$ $a.u.$ polarizabilities. Vector polarizability happen to be below the detection limit of the experimental setup.

## VII. ACKNOWLEDGMENTS

This work was supported by Rosatom in the framework of the Roadmap for Quantum computing (Contract No. 868-1.3-15/15-2021 dated October 5, 2021).